\documentclass[12pt]{article}
\usepackage{graphicx}
\usepackage{color}
\usepackage{cite}
\newcommand{\abs}[1]{\left| #1\right|}
\newcommand{\xrm}[1]{{\textstyle \mbox{\rm #1}}}
\newcommand{\fnd}[2]{\frac{\textstyle #1}{\textstyle #2}}

\newcommand{\bm}[1]{\mbox{\boldmath $#1$}}

\newcommand{\bracket}[2]{\mbox{$\left\langle #1\left| #2\right.\right\rangle$}}
\newcommand{\ket}[1]{\mbox{$\left| #1\right\rangle$}}
\newcommand{\braket}[3]{\mbox{$\left\langle #1\left|
#2\right| #3\right\rangle$}}
\def\chie{\mbox{\raisebox{0.5ex}{$\chi$}}}
\newcommand{\Clebsch}[6]{\mbox{$C\begin{array}{ccc} \!\!\! #1 & \!\!\!\! #2 &
\!\!\!\! #3\\ \!\!\! #4 & \!\!\!\! #5 & \!\!\!\! #6\end{array}$}}
\begin{document} \baselineskip .7cm
\title{\bf Coupling constants and transition potentials
for hadronic decay modes of a meson}
\author{
E. van Beveren$^{1}$,\\
{\normalsize\it Institute for Theoretical Physics, University of Nijmegen}\\
{\normalsize\it NL-6525 ED, Nijmegen, The Netherlands}
}
\footnotetext[1]{present address:
Departamento de F\'{\i}sica, Universidade de Coimbra, P-3004-516 Portugal\\
e-mail: eef@teor.fis.uc.pt\\
url: http://cft.fis.uc.pt/eef}

\date{preprint, 1 August 1983\\ - \\
published in \\ Zeitschrift f\"{u}r Physik {\bf C} - Particles and Fields
{\bf 21}, 291-297 (1984)}
\maketitle

\begin{abstract}
Within the independent-harmonic-oscillator model for quarks
inside a hadron, a rigorous method is presented for the calculation of
coupling constants and transition potentials for hadronic decay, as
needed in a multi-channel description of mesons.
\end{abstract}
\clearpage

\section{Introduction}

In several publications
\cite{PRD17p3090,PRL50p1181,PRD29p110,KAZIMIERZ83p257}
it is discussed that hadronic decay cannot be ignored in hadron models.
As a consequence it is essential
for the description of mesons (and baryons) to know the allowed decay
modes and their relative strengths.

The use of independent harmonic
oscillators for the calculation of the angular-momentum part of
rearrangement matrix elements, has been demonstrated in a previous paper
\cite{ZPC17p135}.
There it was shown that the difficulties, which arise in naive
recoupling schemes \cite{NPB10p521}
when for a decay process the decay products have
internal angular excitations, can easily be handled once one introduces
explicit wave functions for the partons involved.
This fact has already been elaborated in
\cite{PRD8p2223}
and \cite{AP124p61}
and successfully applied to the
calculation of branching ratios for the decay of mesons.
Here we want
to use this as a method for determining the coupling constants and
transition potentials in a multichannel meson-meson scattering model
\cite{PRD21p772,PRD27p1527}.

There is, however, in principle no limit on the number of
possible decay modes in the schemes of
Refs.~\cite{PRD8p2223,AP124p61}.
For this reason we prefer the
rearrangement method of \cite{ZPC17p135}.
Essentially, when one wants to calculate all
possible decay channels of a given initial state,
then the harmonic-oscillator treatment has the advantage
that there is only finite number of possible decay channels involved,
which in a certain sense is complete.
As a consequence, one can always easily verify
whether all possibilities are taken into account.

In this paper we will calculate explicitly the matrix elements for the
decay processes:
\begin{displaymath}
\xrm{meson}\;\to\;\xrm{meson}+\xrm{meson}
\end{displaymath}
(Sects.~\ref{wave_functions},~\ref{MM_final_state},~\ref{Matrix_Elements}).
Of course, the Clebsch-Gordannery is straightforward,
but it is useful to have the complete formulas written somewhere.
Moreover, the here presented strategy is not totally trivial.
We will see that in order to obtain
the proper selection rules for the decay products, we only need to
impose Fermi statistics on the quarks and the antiquarks.
$G$ parity is then automatically obeyed.
In Sect.~\ref{results} some of the resulting coupling constants
are compared to the data.
The construction of a transition potential which can be used in models
\cite{PRD21p772,PRD27p1527}
for the description of hadronic decay, is given in
Sect.~\ref{Transition_Potential}.
\clearpage

\section{The meson just before decay}
\label{wave_functions}

\begin{figure}[htbp]
\begin{center}
\begin{picture}(330,210)(-80,0)
\put(-26,155){\makebox(0,0)[rc]{\large
\bm{q\left( a,\mu_{1}\right)}}}
\put(186,180){\makebox(0,0)[lc]{\large
\bm{\bar{q}\left( b,\mu_{2}\right)}}}
\put(166,80){\makebox(0,0)[lc]{\large
\bm{q\left(\mu_{3}\right)}}}
\put(14,20){\makebox(0,0)[rc]{\large
\bm{\bar{q}\left(\mu_{4}\right)}}}
\put(-10,172){\makebox(0,0)[bc]{\large\bf 1}}
\put(170,197){\makebox(0,0)[bc]{\large\bf 2}}
\put(150,63){\makebox(0,0)[tc]{\large\bf 3}}
\put(30,3){\makebox(0,0)[tc]{\large\bf 4}}
\put(83,164.5){\makebox(0,0)[lt]{\bf CM}}
\put(85,50){\makebox(0,0)[rb]{\bf CM}}
\put(90,114){\makebox(0,0)[lc]{\large\bm{\vec{r}_{12,34}}}}
\put(80,178.5){\makebox(0,0)
{\rotatebox{7.91}{\bf original pair}}}
\put(145,167){\makebox(0,0)
{\rotatebox{7.91}{\large\bm{\vec{r}_{12}}}}}
\put(90,39){\makebox(0,0)
{\rotatebox{26.57}{\bf \bm{{^{3}P}_{0}} pair}}}
\put(45,39){\makebox(0,0)
{\rotatebox{26.57}{\large\bm{\vec{r}_{34}}}}}
\end{picture}
\end{center}
\caption[]{\small
The four-quark system of the original pair and the newly created
${^{3}P}_{0}$ pair, just before decay.}
\label{4BeforeDecay}
\end{figure}
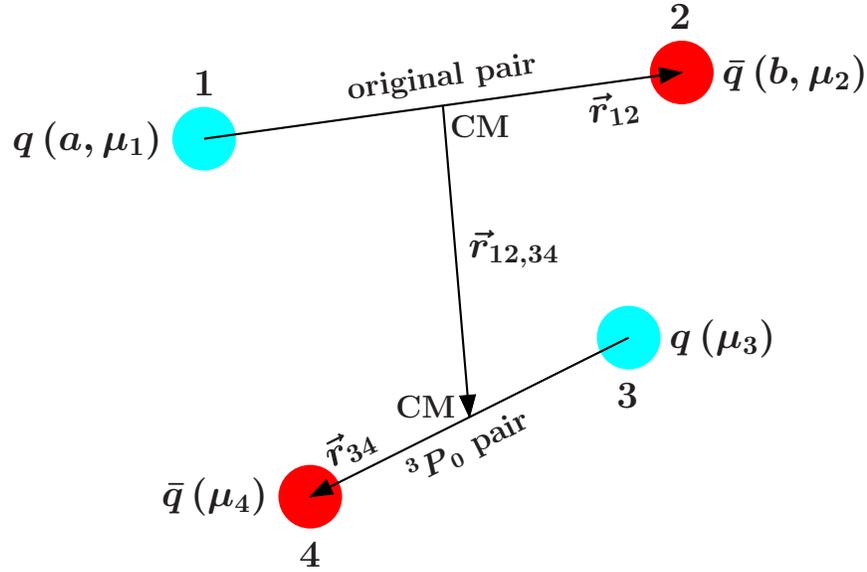

Just before decay a meson is assumed to consist of
four partons in the ${^{3}P_{0}}$ model
(see also Fig.~\ref{4BeforeDecay}):
\begin{enumerate}
\item
the original quark-antiquark ($q\bar{q}$) pair which carries all the
quantum numbers of the decaying meson,
\item
the newly created $q\bar{q}$ pair with the quantum numbers of the vacuum
($J^{PC}=0^{++}$).
\end{enumerate}

First we introduce a quarkspinor which describes a quark with
flavorindex $a$ (here the number of flavors is restricted to three; the
generalization to more flavors is straightforward),
colorindex $\alpha$ and spin magnetic quantum number $\mu$ by
\begin{equation}
q^{a}_{\alpha}\left(\mu\right)
\;\;\; .
\label{quarkwf}
\end{equation}
For an antiquark we use similarly the notation
\begin{equation}
\bar{q}^{\,\alpha}_{a}\left(\mu\right)
\;\;\; .
\label{antiquarkwf}
\end{equation}
\clearpage

The wave function describing the system of four partons
which compose a decaying meson, is the product of two parts:
\begin{enumerate}
\item
The wave function of the original $q\bar{q}$ pair,
which consists of a color-singlet quark-antiquark pair
with (flavor, spin) indices ($a$, $\mu_{1}$) for the quark
and ($b$, $\mu_{2}$) for the antiquark
\begin{eqnarray}
\lefteqn{
\chie_\xrm{\scriptsize meson}\left( J,J_{z};\ell,s,n;a,b;\bm{r}_{12}\right)
\; =}
\label{original_meson}\\ [10pt] & &
\sum_{\scriptsize
\begin{array}{c}\mu_{\ell},\mu_{s}\\ \mu_{1},\mu_{2}\end{array}}\!\!\!
\Clebsch{\ell}{s}{J}{\mu_{\ell}}{\mu_{s}}{J_{z}}\,
\Clebsch{\frac{1}{2}}{\frac{1}{2}}{s}{\mu_{1}}{\mu_{2}}{\mu_{s}}\,
\phi_{\, n,\ell,\mu_{\ell}}\left(\bm{r}_{12}\right)\,
\frac{1}{\sqrt{3}}\,
q^{a}_{\alpha}\left(\mu_{1}\right)
\bar{q}^{\,\alpha}_{a}\left(\mu_{2}\right)
\;\;\; .
\nonumber
\end{eqnarray}
Here and in the following we adopt the standard summation
convention for repeated indices.
The antiquark is at relative position $\bm{r}_{12}$ with respect to
the quark.
This system has relative angular momentum $\ell$,
total spin $s$,
total angular momentum $J$ ($z$ component: $J_{z}$)
and radial excitation of the spatial relative motion $n$.
\item
Similarly the newly created color-singlet and flavor-singlet
${^{3}P_{0}}$ pair is described by
\begin{equation}
\chie_{{^{3}P_{0}}}\left(\bm{r}_{34}\right)
\; =\!\!
\sum_{\mu ,\mu_{3},\mu_{4}}\!
\Clebsch{1}{1}{0}{\mu}{-\mu}{0}\,
\Clebsch{\frac{1}{2}}{\frac{1}{2}}{1}{\mu_{3}}{\mu_{4}}{-\mu}\,
\phi_{\, 0,1,\mu}\left(\bm{r}_{34}\right)\,
\frac{1}{3}\,
q^{c}_{\beta}\left(\mu_{3}\right)
\bar{q}^{\,\beta}_{c}\left(\mu_{4}\right)
\;\;\; .
\label{3P0pair}
\end{equation}
where we have taken the lowest radial excitation in the spatial part.
The precise definition of the harmonic oscillator wave functions
$\phi_{\, n,\ell,m}\left(\bm{r}\right)$ can
be found in \cite{ZPC17p135}.
\end{enumerate}

The total wave function for the four particles must be antisymmetric
with respect to the interchange of either two quarks or two antiquarks.
For this purpose we define the exchange operator $P^{ij}$,
which operator interchanges partons $i$ and $j$.
This way we obtain for the wave function for a meson just before decay
\begin{equation}
\left( 1-P^{13}-P^{24}+P^{13}P^{24}\right)\,
\ket{M+{^{3}P_{0}}\, ;\, J,J_{z};\ell,s,n;a,b;
\bm{r}_{12},\bm{r}_{34},\bm{r}_{12,34}}
\;\;\; ,
\label{BeforeDecay}
\end{equation}
where
\begin{eqnarray}
\lefteqn{
\ket{M+{^{3}P_{0}}\, ;\, J,J_{z};\ell,s,n;a,b;
\bm{r}_{12},\bm{r}_{34},\bm{r}_{12,34}}
\; =}
\nonumber\\ [10pt] & & \;\;\;\;\;\;\; =\;
\chie_\xrm{\scriptsize meson}\left( J,J_{z};\ell,s,n;a,b;\bm{r}_{12}\right)\,
\chie_{{^{3}P_{0}}}\left(\bm{r}_{34}\right)\,
\phi_{\, 0,0,0}\left(\bm{r}_{12,34}\right)
\;\;\; .
\label{KetBefore}
\end{eqnarray}
In Eq.~(\ref{KetBefore}),
the relative motion of the two $q\bar{q}$ pairs,
which have relative position $\bm{r}_{12,34}$, is assumed to have the
ground state quantum numbers.
So far we have not taken into account normalization factors,
but we will come back to this point in Sect.~\ref{Matrix_Elements}.
\clearpage

\section{The meson+meson final state}
\label{MM_final_state}

\begin{figure}[htbp]
\begin{center}
\begin{picture}(330,210)(-80,0)
\put(-26,155){\makebox(0,0)[rc]{\large
\bm{q\left( a',{\mu '}_{1}\right)}}}
\put(186,180){\makebox(0,0)[lc]{\large
\bm{\bar{q}\left( b',{\mu '}_{2}\right)}}}
\put(166,80){\makebox(0,0)[lc]{\large
\bm{q\left( c',{\mu '}_{3}\right)}}}
\put(14,20){\makebox(0,0)[rc]{\large
\bm{\bar{q}\left( d',{\mu '}_{4}\right)}}}
\put(-10,172){\makebox(0,0)[bc]{\large\bf 1}}
\put(170,197){\makebox(0,0)[bc]{\large\bf 2}}
\put(150,63){\makebox(0,0)[tc]{\large\bf 3}}
\put(30,3){\makebox(0,0)[tc]{\large\bf 4}}
\put(110,127){\makebox(0,0)
{\rotatebox{15.82}{\large\bm{\vec{r}_{14,32}}}}}
\put(-1,87.5){\makebox(0,0)
{\rotatebox{-73.5}{\bf meson \bm{a}}}}
\put(155,159){\makebox(0,0)
{\rotatebox{78.69}{\large\bm{\vec{r}_{32}}}}}
\put(171,130){\makebox(0,0)
{\rotatebox{78.69}{\bf meson \bm{b}}}}
\put(32,50){\makebox(0,0)
{\rotatebox{-73.5}{\large\bm{\vec{r}_{14}}}}}
\end{picture}
\end{center}
\caption[]{\small
The four-quark system of the original pair and the newly created
${^{3}P}_{0}$ pair, just before decay.}
\label{4AfterDecay}
\end{figure}
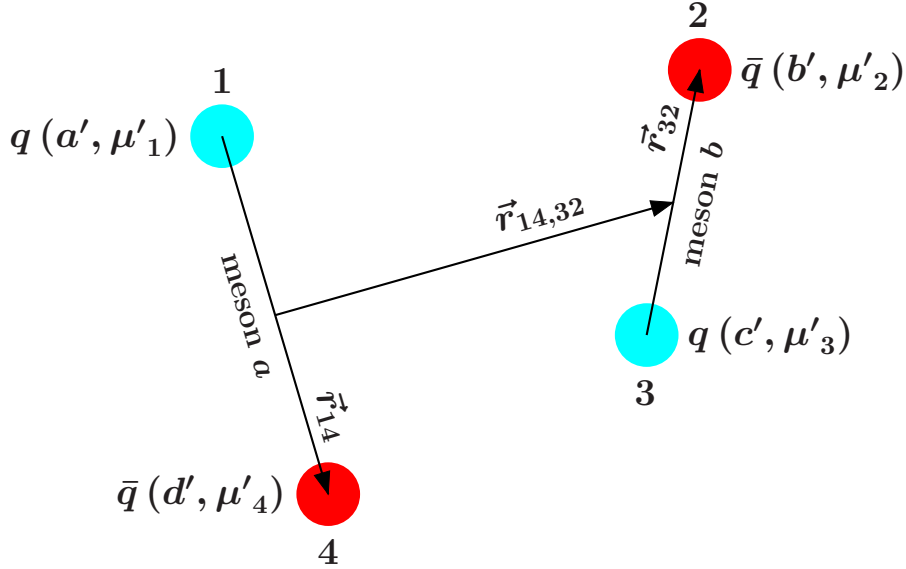

When, out of the two $q\bar{q}$ pairs,
two new mesons are formed, we have a system
of two distinct objects, each of which consisting of a color singlet
$q\bar{q}$ pair (Fig.~\ref{4AfterDecay}).
We will not treat here possible octet $q\bar{q}$ pairs.
The quantum numbers (total angular momentum, orbital angular momentum,
spin and radial orbital excitation) of the mesons are given by
$\left( J_{a},\ell_{a},s_{a},n_{a}\right)$ for meson 1
and by $\left( J_{b},\ell_{b},s_{b},n_{b}\right)$ for meson 2,
whereas their flavor indices are given by
$(a', d')$ and $(c', b')$ respectively.

The wave function must be symmetric under the interchange of the two
mesons, or in terms of the quarks, symmetric under the simultaneous
interchange of quarks and antiquarks, {\it i.e.}
\begin{equation}
\left( 1+P^{13}P^{24}\right)\,
\ket{M+M\, ;\, J,J_{z};\, J_{c},\ell_{c},n_{c};\,
J_{a},\ell_{a},s_{a},n_{a};\, J_{b},\ell_{b},s_{b},n_{b};\,
a',d',c',b';\,
\bm{r}_{14},\bm{r}_{32},\bm{r}_{14,32}}
\;\;\; ,
\label{AfterDecay}
\end{equation}
where
\begin{eqnarray}
\lefteqn{
\ket{M+M\, ;\, J,J_{z};\, J_{c},\ell_{c},n_{c};\,
J_{a},\ell_{a},s_{a},n_{a};\, J_{b},\ell_{b},s_{b},n_{b};\,
a',d',c',b';\,
\bm{r}_{14},\bm{r}_{32},\bm{r}_{14,32}}
\; =}
\nonumber\\ [10pt] & & \!\!\!\!\!\!\!\! =\!\!\!\!
\sum_{\scriptsize
\begin{array}{c}m_{a},\mu_{a}\\ m_{b},\mu_{b}\\
\;\; m_{c}\end{array}}\!\!\!
\Clebsch{J_{c}}{\ell_{c}}{J}{M_{c}}{m_{c}}{J_{z}}\,
\Clebsch{J_{a}}{J_{b}}{J_{c}}{M_{a}}{M_{b}}{M_{c}}\,
\Clebsch{\ell_{a}}{s_{a}}{J_{a}}{m_{a}}{\mu_{a}}{M_{a}}\,
\Clebsch{\ell_{b}}{s_{b}}{J_{b}}{m_{b}}{\mu_{b}}{M_{b}}\,
\label{KetAfter}\\ & &
\phi_{\, n_{a},\ell_{a},m_{a}}\left(\bm{r}_{14}\right)\,
\phi_{\, n_{b},\ell_{b},m_{b}}\left(\bm{r}_{32}\right)\,
\phi_{\, n_{c},\ell_{c},m_{c}}\left(\bm{r}_{14,32}\right)
\!\!\!\sum_{\scriptsize
\begin{array}{c}{\mu '}_{1},{\mu '}_{2}\\
{\mu '}_{3},{\mu '}_{4}\end{array}}\!\!\!
\frac{1}{3}\,
q^{a'}_{\alpha}\left({\mu '}_{1}\right)
\bar{q}^{\,\alpha}_{d'}\left({\mu '}_{4}\right)
q^{c'}_{\beta}\left({\mu '}_{3}\right)
\bar{q}^{\,\beta}_{b'}\left({\mu '}_{2}\right)
\; .
\nonumber
\end{eqnarray}
\clearpage

\section{The matrix elements}
\label{Matrix_Elements}

We obtained now two wave functions of the following structure:
\begin{enumerate}
\item
before decay (see Eq.~\ref{BeforeDecay} and Fig.~\ref{4BeforeDecay}):
\begin{equation}
\left( 1-P^{13}-P^{24}+P^{13}P^{24}\right)\,
\ket{M+{^{3}P_{0}}}
\;\;\; ,
\label{gBeforeDecay}
\end{equation}
\item
after decay (see Eq.~\ref{AfterDecay} and Fig.~\ref{4AfterDecay}):
\begin{equation}
\left( 1+P^{13}P^{24}\right)\,\ket{M+M}
\;\;\; ,
\label{gAfterDecay}
\end{equation}
\end{enumerate}
We demand the wave function expressed by the ket $\ket{M+M}$
in Eq.~(\ref{gAfterDecay}) to be such that the
following matrix elements vanish,
\begin{equation}
\braket{M+{^{3}P_{0}}}{P^{13}}{M+M}\; =\;
\braket{M+{^{3}P_{0}}}{P^{24}}{M+M}\; =\; 0
\;\;\; ,
\label{vanish}
\end{equation}
{\it i.e.},
when we interchange only the quarks or only the antiquarks,
then the wave function must have such a structure that
the result is orthogonal to the states we consider.
This has to do with the fact that nature
seems to know which quark and antiquark belong to each other.
As we may see, it also expresses the fact that nature
obeys the OZI rule, or satisfies a kind of string idea.
We ignore here the fact that, of course,
in nature the relation (\ref{vanish})
will only be approximately satisfied,
because the meson wave functions have some spreading.
The here allowed processes are (OZI allowed)
\begin{equation}
a\bar{b}\,\to\, a\bar{q}\, +\, q\bar{b}
\;\;\;\xrm{or}\;\;\;
a\bar{b}\,\to\, q\bar{b}\, +\, a\bar{q}
\;\;\; ,
\label{OZIallowed}
\end{equation}
the non-allowed processes are (OZI forbidden)
\begin{equation}
a\bar{b}\,\to\, a\bar{b}\, +\, q\bar{q}
\;\;\;\xrm{or}\;\;\;
a\bar{b}\,\to\, q\bar{q}\, +\, a\bar{b}
\;\;\; .
\label{forbidden}
\end{equation}

The matrix-element
\begin{equation}
{\cal M}\; =\;
\braket{M+M}{\left( 1+P^{13}P^{24}\right)^{\dagger}
\left( 1-P^{13}-P^{24}+P^{13}P^{24}\right)}{M+{^{3}P_{0}}}
\;\;\; ,
\label{Mdef}
\end{equation}
involves the product of exchange operators
\begin{equation}
\left( 1+P^{13}P^{24}\right)^{\dagger}
\left( 1-P^{13}-P^{24}+P^{13}P^{24}\right)
\; =\;
2\,\left( 1-P^{13}-P^{24}+P^{13}P^{24}\right)
\;\;\; .
\label{Pproduct}
\end{equation}
When we combine Eq.~(\ref{Pproduct})
with the conditions (\ref{vanish}),
then, besides a factor 2,
we obtain for the transition matrix element (\ref{Mdef}):
\begin{eqnarray}
{\cal M} & = &
\braket{M+M\, ;\,\bm{r}_{14},\bm{r}_{32},\bm{r}_{14,32}}
{\left( 1+P^{13}P^{24}\right)}
{M+{^{3}P_{0}}\, ;\,\bm{r}_{12},\bm{r}_{34},\bm{r}_{12,34}}
\nonumber\\ [10pt] & = &
\braket{M+M\, ;\,\bm{r}_{12},\bm{r}_{34},\bm{r}_{12,34}}
{\left( P^{13}+P^{24}\right)}
{M+{^{3}P_{0}}\, ;\,\bm{r}_{12},\bm{r}_{34},\bm{r}_{12,34}}
\;\;\; .
\label{Mdef1}
\end{eqnarray}

Since the normalization constants are not yet relevant,
we prefer to take formula (\ref{Mdef1})
for the definitions of the wave functions, rather than the
wave functions defined
in Eqs.~(\ref{BeforeDecay}) and (\ref{AfterDecay}),
{\it i.e.}
\begin{enumerate}
\item
before decay:
\begin{equation}
{\cal N}_{1}
\left[ J,J_{z};\,\ell ,s,n;\, (a,b)\right]\,
\left( P^{13}+P^{24}\right)\,
\ket{M+{^{3}P_{0}}}
\;\;\; ,
\label{rBeforeDecay}
\end{equation}
\item
after decay:
\begin{equation}
{\cal N}_{2}
\left[ J,J_{z};\, J_{c},\ell_{c},n_{c};\,
J_{a},\ell_{a},s_{a},n_{a};\, J_{b},\ell_{b},s_{b},n_{b};\,
(a',d'),(c',b')\right]\,\ket{M+M}
\;\;\; .
\label{rAfterDecay}
\end{equation}
\end{enumerate}

When we use the harmonic oscillator wave functions
which are defined in \cite{ZPC17p135},
then we obtain for the normalization constants in
Eqs.~(\ref{rBeforeDecay}) and (\ref{rAfterDecay}),
\begin{equation}
{\cal N}_{1}
\left[ J,J_{z};\,\ell ,s,n;\, (a,b)\right]\; =\;
\left[ 1+
\delta_{J,0}\,\delta_{J_{z},0}\,\delta_{\ell ,1}\,\delta_{s,1}
\,\delta_{n,0}\,\abs{\bracket{a\bar{b}}{\{ 1\}}}^{2}\right]^{-\frac{1}{2}}
\;\;\; ,
\label{N1}
\end{equation}
which is only different from 1 when
the original quark-antiquark pair $(a,b)$
is in a flavor-singlet ${^{3}P_{0}}$ ground state.
$P^{13}$ interchanges the quarks and $P^{24}$ the antiquarks.
The bracket $\bracket{a\bar{b}}{\{ 1\}}$ represents
the coefficient of the $SU(3)$-singlet component in the full
flavor wave function of the initial meson.

For ${\cal N}_{2}$ we find
\begin{equation}
{\cal N}_{2}
\left[ J,J_{z};\, J_{c},\ell_{c},n_{c};\,
J_{a},\ell_{a},s_{a},n_{a};\, J_{b},\ell_{b},s_{b},n_{b};\,
(a',d'),(c',b')\right]
\; =\; 1
\;\;\; .
\label{N2}
\end{equation}

\begin{figure}[htbp]
\begin{center}
\begin{tabular}{c}
\includegraphics[height=240pt]{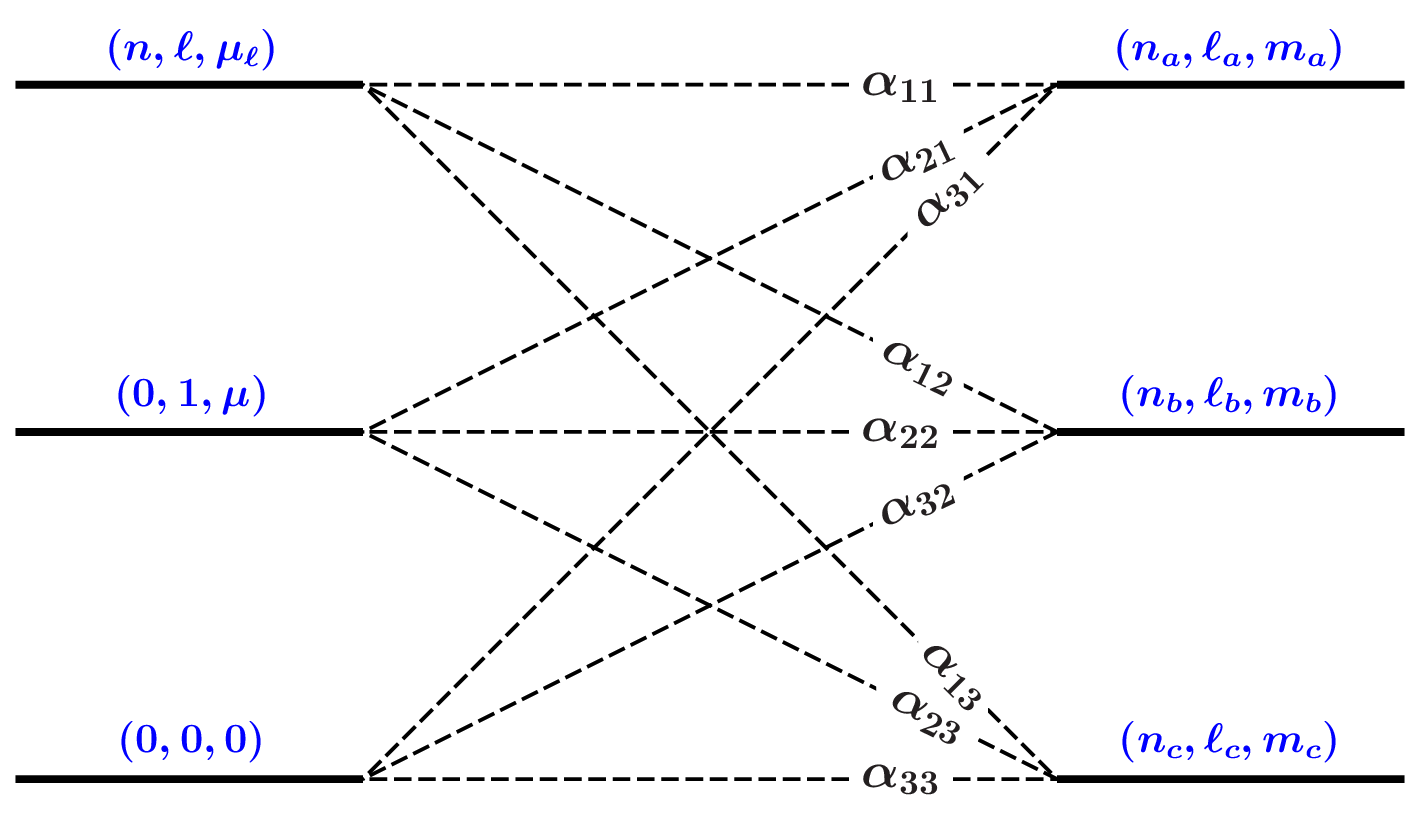}
\end{tabular}
\end{center}
\caption[]{\small Diagrammatic representation of the spatial parts
of formula (\ref{Mdef1}).}
\label{MtoMM}
\end{figure}

The spatial parts of the matrix element (\ref{Mdef1})
consist out of the diagrams as represented in Fig.~\ref{MtoMM}
(see Ref.~\cite{ZPC17p135} for the definitions).
The internal lines of diagram (\ref{MtoMM}) are function of
the matrix elements $\alpha_{11}$, ..., $\alpha_{33}$
of a $3\times 3$ matrix $\alpha$.
When we restrict ourselves to the case of equal quark masses,
then $\alpha$ is for the first term of (\ref{Mdef1}) given by
\begin{equation}
\left(
\begin{array}{l}
\bm{r}_{32}\\ [10pt]
\bm{r}_{14}\\ [10pt]
\bm{r}_{32,14}
\end{array}
\right)\; =\;
\left(\begin{array}{ccc}
\frac{1}{2} & \frac{1}{2} &  \sqrt{\frac{1}{2}}\\ [10pt]
\frac{1}{2} & \frac{1}{2} & -\sqrt{\frac{1}{2}}\\ [10pt]
\sqrt{\frac{1}{2}} & -\sqrt{\frac{1}{2}} & 0\end{array}\right)
\left(
\begin{array}{l}
\bm{r}_{12}\\ [10pt]
\bm{r}_{34}\\ [10pt]
\bm{r}_{12,34}
\end{array}
\right)
\;\;\; ,
\label{alphaP13}
\end{equation}
and for the second term by
\begin{equation}
\left(
\begin{array}{l}
\bm{r}_{14}\\ [10pt]
\bm{r}_{32}\\ [10pt]
\bm{r}_{14,32}
\end{array}
\right)\; =\;
\left(\begin{array}{ccc}
\frac{1}{2} & \frac{1}{2} & -\sqrt{\frac{1}{2}}\\ [10pt]
\frac{1}{2} & \frac{1}{2} & \sqrt{\frac{1}{2}}\\ [10pt]
-\sqrt{\frac{1}{2}} & \sqrt{\frac{1}{2}} & 0\end{array}\right)
\left(
\begin{array}{l}
\bm{r}_{12}\\ [10pt]
\bm{r}_{34}\\ [10pt]
\bm{r}_{12,34}
\end{array}
\right)
\;\;\; ,
\label{alphaP24}
\end{equation}

We denote the respective results of the diagram of Fig.~\ref{MtoMM}
for the $\alpha$'s defined in Eqs.~(\ref{alphaP13})
and (\ref{alphaP24}), respectively by
\begin{equation}
d_{1}\left(
n,\ell ,\mu_{1};\,\mu ;\,
n_{a},\ell_{a},m_{a};\,
n_{b},\ell_{b},m_{b};\,
n_{c},\ell_{c},m_{c}\right)
\;\;\; ,
\label{d1}
\end{equation}
and
\begin{equation}
d_{2}\left(
n,\ell ,\mu_{1};\,\mu ;\,
n_{a},\ell_{a},m_{a};\,
n_{b},\ell_{b},m_{b};\,
n_{c},\ell_{c},m_{c}\right)
\;\;\; .
\label{d2}
\end{equation}
With these definitions we obtain for the matrix elements (\ref{Mdef1}):
\begin{eqnarray}
{\cal M} & = &
{\cal N}_{1}\left[ J,J_{z};\,\ell ,s,n;\, (a,b)\right]
\sum_{\scriptsize\begin{array}{c}
m_{a},\mu_{a},m_{b},\mu_{b}\\
m_{c},\mu_{\ell},\mu_{s},\mu_{1},\mu_{2}\\
m_{1},m_{2},m_{3},m_{4}
\end{array}}\!\!\!
\Clebsch{J_{c}}{\ell_{c}}{J}{M_{c}}{m_{c}}{J_{z}}\,
\Clebsch{J_{a}}{J_{b}}{J_{c}}{M_{a}}{M_{b}}{M_{c}}\,
\label{Mfinal}\\ [10pt] & &
\Clebsch{\ell_{a}}{s_{a}}{J_{a}}{m_{a}}{\mu_{a}}{M_{a}}\,
\Clebsch{\ell_{b}}{s_{b}}{J_{b}}{m_{b}}{\mu_{b}}{M_{b}}\,
\Clebsch{\ell}{s}{J}{\mu_{\ell}}{\mu_{s}}{J_{z}}\,
\Clebsch{\, 1}{1}{0}{\mu_{1}}{\mu_{2}}{0}\,
\Clebsch{\frac{1}{2}}{\frac{1}{2}}{s_{a}}{m_{1}}{m_{2}}{\mu_{a}}\,
\Clebsch{\frac{1}{2}}{\frac{1}{2}}{s_{b}}{m_{3}}{m_{4}}{\mu_{b}}
\nonumber\\ [10pt] & &
\left\{
d_{1}\left(
n,\ell ,\mu_{1};\,\mu ;\,
n_{a},\ell_{a},m_{a};\,
n_{b},\ell_{b},m_{b};\,
n_{c},\ell_{c},m_{c}\right)\,
\Clebsch{\frac{1}{2}}{\frac{1}{2}}{s}{m_{3}}{m_{2}}{\mu_{s}}\,
\Clebsch{\frac{1}{2}}{\frac{1}{2}}{1}{m_{1}}{m_{4}}{\mu_{2}}\,
\delta_{a'b'}\,\delta_{d'b}\,\delta_{c'a}\; +\right.
\nonumber\\ [10pt] & &
\left. +\;
d_{2}\left(
n,\ell ,\mu_{1};\,\mu ;\,
n_{a},\ell_{a},m_{a};\,
n_{b},\ell_{b},m_{b};\,
n_{c},\ell_{c},m_{c}\right)\,
\Clebsch{\frac{1}{2}}{\frac{1}{2}}{s}{m_{1}}{m_{4}}{\mu_{s}}\,
\Clebsch{\frac{1}{2}}{\frac{1}{2}}{1}{m_{3}}{m_{2}}{\mu_{2}}
\delta_{a'a}\,\delta_{d'c'}\,\delta_{b'b}\right\}
\; .
\nonumber
\end{eqnarray}
In Sect.~\ref{wave_functions}
we smuggled two assumptions into the procedure, which had to
do with the excitations of the $^{3}P_{0}$-pair.
This resulted in the four zero's in diagram (\ref{MtoMM}).
The generalization to other quantum
numbers is straightforward but will not be treated here.
\clearpage

\section{Results and comparison with experiment}
\label{results}

In this section we study the results of expression (\ref{Mfinal})
for some quark-antiquark systems.
First we introduce the nomenclature which has been used here for mesons.
In Table~\ref{NoMen}, we give the precise meaning of the symbols
which we use in Table~\ref{Vertices}.
\begin{table}[htbp]
\begin{center}
\begin{tabular}{||l|l|c|c||}
\hline\hline & & & \\ [-5pt]
meson & $q\bar{q}$ & $n{^{2S+1}L}_{J}$ & $J^{PC}$\\
& & & \\ [-5pt]
\hline\hline & & & \\ [-5pt]
pseudoscalars & ground state & 1$\,{^{1}S}_{0}$ & $0^{-+}$\\
& first radial excitation & 2$\,{^{1}S}_{0}$ & ${0'}^{-+}$\\
& & & \\ [-5pt]
\hline & & & \\ [-5pt]
vectors & ground state & 1$\,{^{3}S}_{1}$ & $1^{--}$\\
& first radial excitation & 2$\,{^{3}S}_{1}$ & ${1'}^{--}$\\
& second angular excitation & 1$\,{^{3}D}_{1}$ & ${1''}^{--}$\\
& & & \\ [-5pt]
\hline & & & \\ [-5pt]
scalars & ground state & 1$\,{^{3}P}_{0}$ & $0^{++}$\\
& & & \\ [-5pt]
\hline & & & \\ [-5pt]
axial vectors & ground state ($S=1$) & 1$\,{^{3}P}_{1}$ & $1^{++}$\\
& ground state ($S=0$) & 1$\,{^{1}P}_{1}$ & $1^{+-}$\\
& & & \\ [-5pt]
\hline\hline
\end{tabular}
\end{center}
\caption[]{\small
Nomenclature of quark-antiquark systems.
The columns contain respectively,
the most current denomination for mesons,
the assumed $q\bar{q}$ state,
the $q\bar{q}$-quantum numbers
($n$ is the assumed radial quantum-number,
$S$ represents the total spin,
$L$ the orbital and $J$ total angular momentum),
and the more common quantum numbers
($P$ stands for parity and $C$ for charge conjugation;
excitations are indicated by accents).}
\label{NoMen}
\end{table}
\clearpage

In Table~\ref{Vertices},
all ({\it i.e.} within the above specified assumptions)
possible decay channels for the lowest radial and angular excitations of
pseudoscalar, vector and scalar mesons are given.
\begin{table}[htbp]
\begin{center}
\begin{tabular}{||l|r|r|r||}
\hline\hline & & & \\ [-5pt]
& $0^{-+}$ & $1^{--}$ & $0^{++}$\\
& & & \\ [-5pt]
\hline\hline & & & \\ [-5pt]
$(0^{-+}\, ,0^{-+}\, )$ & &
$-\frac{1}{24}\left( 1{^{1}P}\right)$ &
$\frac{1}{24}\left( 2{^{1}S}\right)$\\ [5pt]
$(0^{-+}\, ,{0'}^{-+}\, )$ & & &
$\frac{1}{48}\left( 1{^{1}S}\right)$\\ [5pt]
$(0^{-+}\, ,1^{--}\, )$ &
$-\frac{1}{4}\left( 1{^{3}P}\right)$ &
$\frac{1}{6}\left( 1{^{3}P}\right)$ & \\ [5pt]
$(0^{-+}\, ,0^{++}\, )$ &
$\frac{1}{8}\left( 1{^{1}S}\right)$ & & \\ [5pt]
$(0^{-+}\, ,1^{++}\, )$ & &
$-\frac{1}{12}\left( 1{^{3}S}\right)$ &
$\frac{1}{6}\left( 1{^{3}P}\right)$\\ [5pt]
$(0^{-+}\, ,1^{+-}\, )$ & &
$\frac{1}{24}\left( 1{^{3}S}\right)$ & \\ [5pt]
$(1^{--}\, ,1^{--}\, )$ &
$\frac{1}{4}\left( 1{^{3}P}\right)$ &
$-\frac{1}{72}\left( 1{^{1}P}\right)$ &
$\frac{1}{72}\left( 2{^{1}S}\right)$ \\ [5pt]
$(1^{--}\, ,1^{--}\, )$ & &
$-\frac{5}{18}\left( 1{^{5}P}\right)$ &
$\frac{5}{18}\left( 1{^{5}D}\right)$ \\ [5pt]
$(1^{--}\, ,{1'}^{--}\, )$ & & &
$\frac{1}{144}\left( 1{^{1}S}\right)$ \\ [5pt]
$(1^{--}\, ,{1''}^{--}\, )$ & & &
$\frac{5}{36}\left( 1{^{1}S}\right)$ \\ [5pt]
$(1^{--}\, ,1^{++}\, )$ &
$-\frac{1}{4}\left( 1{^{1}S}\right)$ &
$\frac{1}{6}\left( 1{^{3}S}\right)$ & \\ [5pt]
$(1^{--}\, ,1^{+-}\, )$ &
$\frac{1}{8}\left( 1{^{1}S}\right)$ &
$-\frac{1}{12}\left( 1{^{3}S}\right)$ &
$\frac{1}{6}\left( 1{^{3}P}\right)$ \\ [5pt]
$(0^{++}\, ,0^{++}\, )$ & & &
$\frac{1}{16}\left( 1{^{1}S}\right)$ \\ [5pt]
$(0^{++}\, ,1^{--}\, )$ & &
$\frac{1}{8}\left( 1{^{3}S}\right)$ & \\ [5pt]
$(1^{++}\, ,1^{++}\, )$ & & &
$\frac{1}{12}\left( 1{^{1}S}\right)$ \\ [5pt]
$(1^{+-}\, ,1^{+-}\, )$ & & &
$\frac{1}{48}\left( 1{^{1}S}\right)$ \\
& & & \\ [-5pt]
\hline\hline
\end{tabular}
\end{center}
\caption[]{\small
Three-meson vertices $\abs{\braket{MM}{V}{M}}^{2}$.
The signs in front of the matrixelements are explained in the text.
The orbital quantum numbers of the meson-meson system
are given between brackets, in the spectroscopic
notation $n{^{2S+1}L}$,
where $n$ is radial quantum number, $S$ the spin
and $L$ the angular momentum.}
\label{Vertices}
\end{table}
Note that the figures in each column of Table~\ref{Vertices} add up to 1.
This expresses the fact that each wave function (\ref{rBeforeDecay})
can be completely decomposed into the basis (\ref{rAfterDecay}),
because both sets are solutions of the same Hamiltonian.

The signs which are given in Table~\ref{Vertices} are the results
of the ratios of the terms stemming from $P^{13}$
and the terms stemming from $P^{24}$ in formula (\ref{Mdef1}).
These signs are essential for conservation of $G$-parity, which
can easily been seen once the $SU(3)$-flavor content is taken into
account.

We are now prepared to test the results of Table~\ref{Vertices}
to the available data.

In general the width of a decaying particle is given
(see {\it e.g.} Ref.~\cite{Pilkuhn}) by
\begin{equation}
\Gamma\; =\;\fnd{k}{4m^{2}}\,\int\, d\cos (\vartheta )\,
\abs{\fnd{T\left( m^{2},\cos (\vartheta )\right)}
{\sqrt{4\pi}}}^{2}
\;\;\; .
\label{width}
\end{equation}
The matrix element $T$ in Eq.~(\ref{width})
includes the calculation of the overlap
between the free meson wave function and the harmonic oscillator basis,
which we have chosen.
In fact we should do
\begin{equation}
\braket{MM}{V}{M}\; =\;\sum_{n,n'}\,
\left\langle MM\left| n'\right.\right\rangle
\braket{n'}{V}{n}
\left\langle n\left| M\right.\right\rangle
\;\;\; ,
\label{matrix_element}
\end{equation}
where $\ket{M}$ is a specific initial state of the
$q\bar{q}$ mesonic system and $\ket{MM}$
represents the final state of the two decay products.
But, since decay
widths can be better calculated from the solutions of the multichannel
Schr\"{o}dinger equation involving the transition potential
(this is done by us in a different program \cite{PRD21p772,PRD27p1527}),
we will not be too rigorous in
demonstrating that the matrix elements (\ref{width}) make sense.
So we assume that
$\left\langle n\left| M\right.\right\rangle$
is only nonzero for one value of $n$,
and that then also the corresponding
$\left\langle MM\left| n'\right.\right\rangle$
is the only nonzero term.
The overlap of the plane waves for
the two meson system and a harmonic oscillator contains a momentum
dependence of the form
\begin{equation}
T\;\propto\; k^{\ell}
\;\;\; .
\label{Tk}
\end{equation}
where $\ell$ is the relative orbital angular momentum of the decay products.
Relation (\ref{Tk}), together with the phase space factor, gives
\begin{equation}
\Gamma\;\propto\; k^{2\ell +1}
\;\;\; .
\label{width1}
\end{equation}
So, we arrive at
\begin{equation}
\Gamma\;\propto\; k^{2\ell +1}\,\abs{\braket{n'}{V}{n}}^{2}
\;\;\; ,
\label{width2}
\end{equation}
where $n$ represents the radial quantum number of the decaying meson.
A more precise form of formula (\ref{width2}) is given by
the imaginary part of the second order
perturbation contribution to the energy within the model
\cite{PRD21p772,PRD27p1527}, as is
given in \cite{ZPC19p275,AP105p318}.
We will, however, take (\ref{width2}) as sufficiently
accurate for our purpose here.
\clearpage

As examples we study the decay widths of
$\epsilon$(1300) (nowadays $f_{0}$(1370)),
$a_{2}$(1320), and $K^{\ast}_{0}$(1430),
which particles have enough two particle decay modes,
to make a comparison of relation (\ref{width2}) to the
available data possible.
The results are given in Table~\ref{branchingratios}.
\begin{table}[htbp]
\begin{center}
\begin{tabular}{||l|l|l|l||}
\hline\hline & & & \\ [-5pt]
processes & Eq.~(\ref{width2}) & PDG80 & PDG04\\
& & & \\ [-5pt]
\hline\hline & & & \\ [-5pt]
$\fnd{\epsilon\to K\bar{K}}{\epsilon\to\pi\pi}$ &
0.22 & 0.11 & 0.12$\pm$0.06 - 0.91$\pm$0.20\\
\hline & & & \\ [-5pt]
$\fnd{a_{2}\to\eta\pi}{a_{2}\to\rho\pi}$ &
0.37 & 0.20 - 0.25 & 0.213$\pm$0.020\\
\hline & & & \\ [-5pt]
$\fnd{K^{\ast}_{0}\to K\pi}{K^{\ast}_{0}\to K^{\ast}\pi}$ &
2.28 & 1.6 - 2.1 & 2.02$\pm$0.14\\
\hline & & & \\ [-5pt]
$\fnd{K^{\ast}_{0}\to K\pi}{K^{\ast}_{0}\to K\rho}$ &
7.76 & 4.3 - 6.1 & 6.7$\pm$1.1\\
\hline & & & \\ [-5pt]
$\fnd{K^{\ast}_{0}\to K\eta}{K^{\ast}_{0}\to K\omega}$ &
2.93 & 0 - 3.7 & - \\
& & & \\ [-5pt]
\hline\hline
\end{tabular}
\end{center}
\caption[]{\small
Some branching ratios of $\epsilon$(1300) (nowadays $f_{0}$(1370)),
$a_{2}$(1320), and $K^{\ast}_{0}$(1430).
Results of expression (\ref{width2}) compared to experiment:
PDG80, C.~Bricman et al., Rev.\ Mod.\ Phys.\ 52, 1 (1980),
and PDG04, S.~Eidelman et al., Phys.\ Lett.\ B592, 1 (2004)
}
\label{branchingratios}
\end{table}
\clearpage

\section{The transition potential}
\label{Transition_Potential}

Inspired by the reasonable results of the comparison of our handwaving
calculations with the data, we will derive in this section a transition
potential for a coupled channel Schr\"{o}dinger equation as described in
\cite{PRD21p772,PRD27p1527}.
In \cite{ZPC17p135} we have shown how we might arrive at a local
approximation of the potential for transitions from a quark-antiquark
channel to a meson-meson channel.
The outline given there will be made more precise in this section.

The potential is given by
\begin{equation}
\braket{MM,x}{V}{M,x}\; =\;\sum_{n,n'}\,
\left\langle MM,x\left| n'\right.\right\rangle
\braket{n'}{V}{n}
\left\langle n\left| M,x\right.\right\rangle
\;\;\; ,
\label{potentialMtoMM}
\end{equation}
The wave functions are in lowest order assumed to be given by
\begin{equation}
\left\langle n\left| M,x\right.\right\rangle\; =\;
\phi_{\, n,\ell ,\mu_{\ell}}\left(\sqrt{\frac{1}{2}m\omega}\, x\right)
\;\;\; ,
\label{meson_wave_function}
\end{equation}
where ($\ell$, $\mu_{\ell}$) are the orbital quantum numbers
of the $q\bar{q}$ mesonic system, where $m$ is the quarkmass
and where $\omega$ is the universal oscillator frequency,
and by
\begin{equation}
\left\langle n'\left| MM,x\right.\right\rangle\; =\;
\phi_{\, n',\ell ',{\mu '}_{\ell}}\left(\sqrt{m\omega}\, x\right)
\;\;\; ,
\label{meson_meson_wave_function}
\end{equation}
where ($\ell '$, ${\mu '}_{\ell}$) are the orbital quantum numbers
of the two meson system.

The matrix elements $\braket{n'}{V}{n}$ are those given in
\cite{ZPC17p135}.
For the lowest lying $q\bar{q}$ systems $n'$ is linearly related to $n$.
\clearpage

\begin{figure}[htbp]
\begin{center}
\begin{tabular}{c}
\includegraphics[height=329pt]
{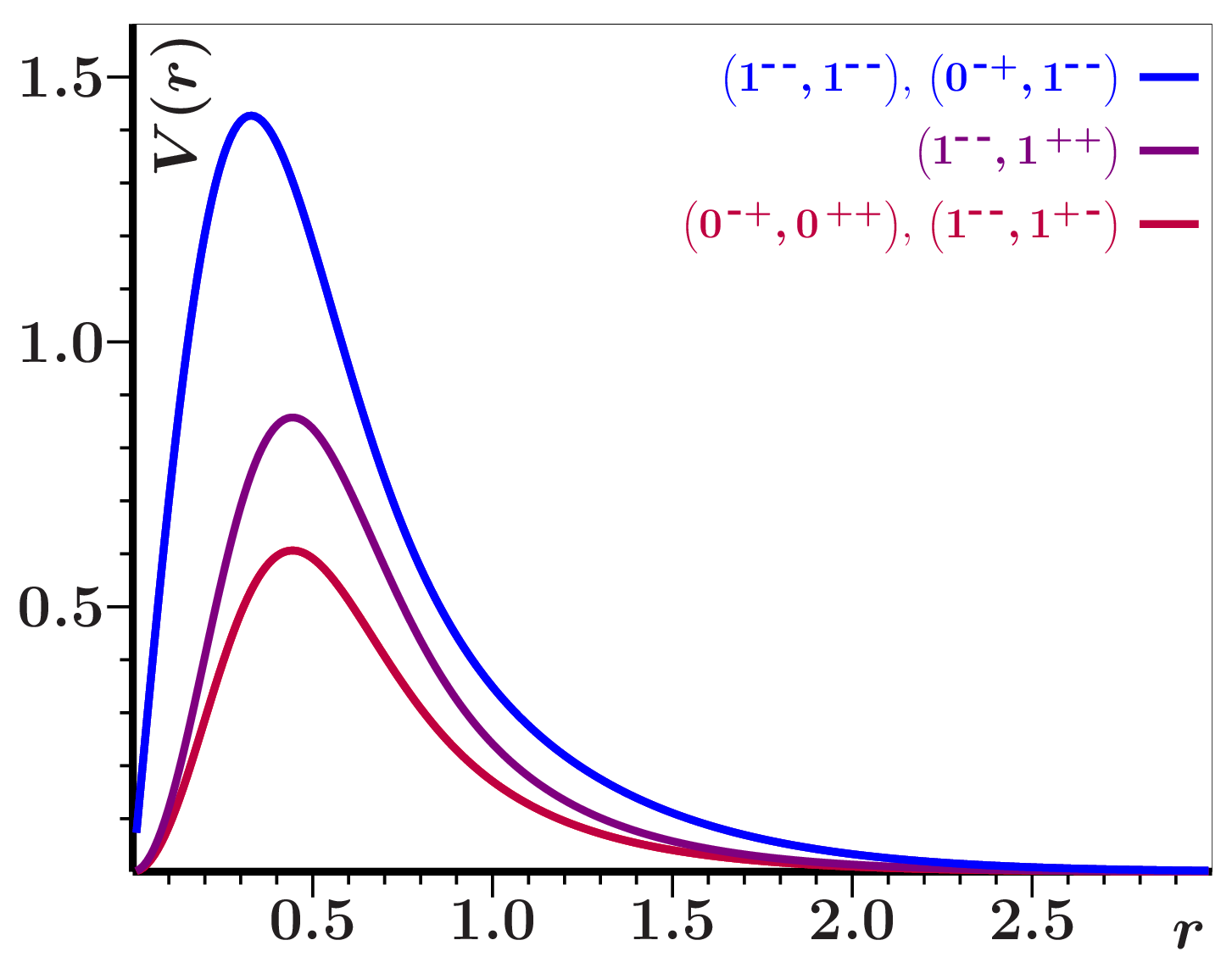}
\end{tabular}
\end{center}
\caption[]{\small
Transition potentials for the decay of pseu\-do\-sca\-lar mesons
($r$ and $V(r)$ are dimensionless).
The curves are determined with expression (\ref{potentialMtoMM})
for the five three-meson vertices of Table~(\ref{Vertices}).\\
The upper curve (blue) corresponds to the pseu\-do\-sca\-lar vertices
with (vector, vector) and (pseu\-do\-sca\-lar, vector)
the middle curve (violet) to the pseu\-do\-sca\-lar vertex with
(vector, positive-$C$-parity axial vector),
the lower curve (red) to the pseu\-do\-sca\-lar vertices
with (pseu\-do\-sca\-lar, scalar) and
(vector, negative-$C$-parity axial vector).}
\label{pseudoscalar}
\end{figure}
In Fig.~\ref{pseudoscalar} we present the results of
Eq.~(\ref{potentialMtoMM}) for the decay processes of
pseudoscalar particles.
Although in Table~\ref{Vertices} five different types of
decay channels for a pseudoscalar meson are shown,
we find that the spatial behavior for two pairs is exactly the same.
We also observe from Fig.~\ref{pseudoscalar}
that the decay processes $P\to PV$ and $P\to VV$ are dominant.
This result is important because it justifies the
procedure of \cite{PRD21p772,PRD27p1527}.
There we have limited us to these channels.
For doing so we now have two arguments: the thresholds are low, the
couplings are largest.
We also find that the form of the here calculated transition
potential is in agreement with the one from \cite{PRD27p1527}.
However, in \cite{PRD27p1527} the
position of the peak is a parameter which can be fitted to the data,
whereas here it is a result of our procedure.
\clearpage

\begin{figure}[htbp]
\begin{center}
\begin{tabular}{c}
\includegraphics[height=329pt]
{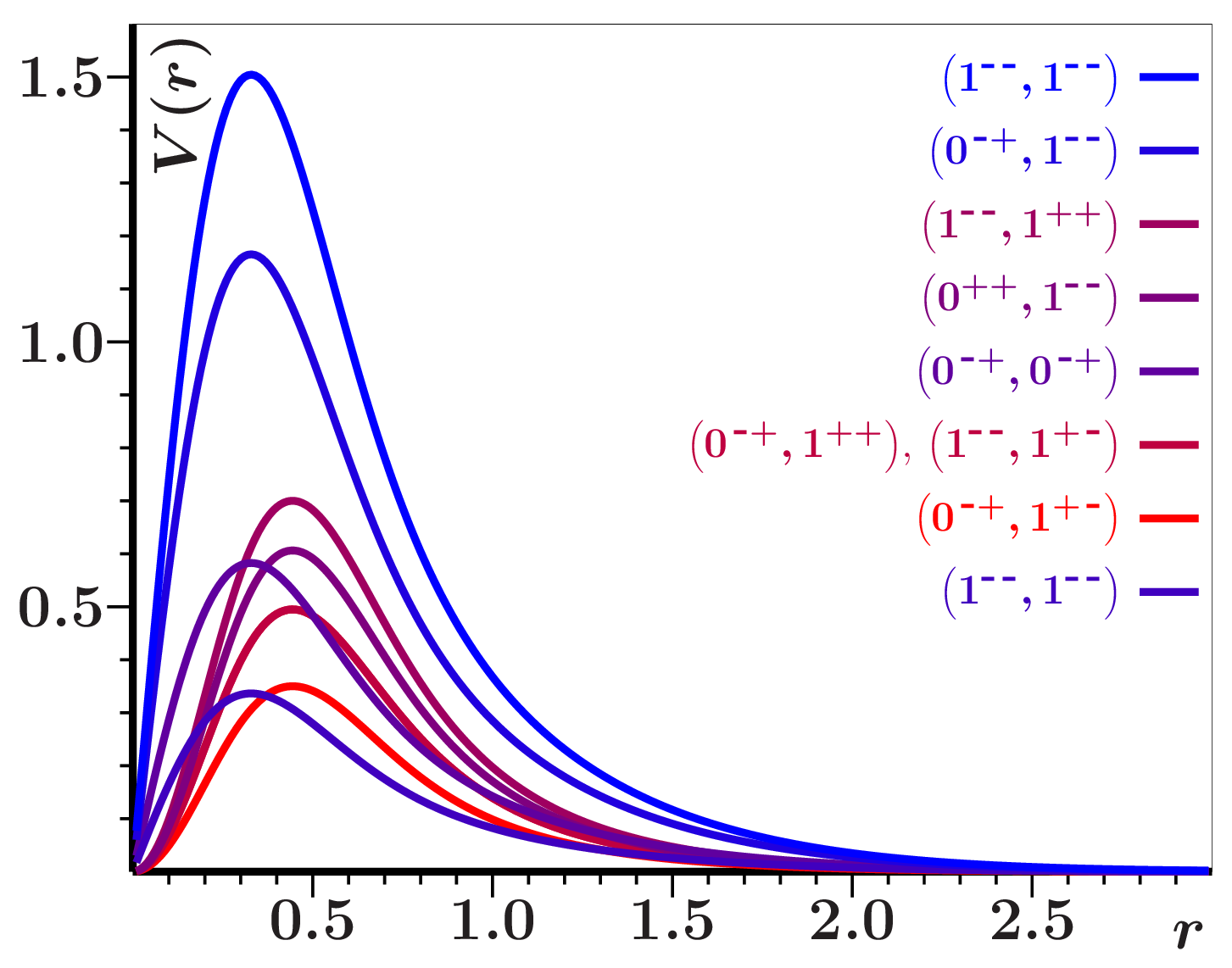}
\end{tabular}
\end{center}
\caption[]{\small
Transition potentials for the decay of vector mesons
($r$ and $V(r)$ are dimensionless).
The curves are determined with expression (\ref{potentialMtoMM})
for the nine three-meson vertices of Table~(\ref{Vertices}).\\
The curves,
from the curve with the highest maximum to the curve with the lowest maximum,
represent the vector vertices with respectively
(vector, vector in $P$ wave, total spin $S=2$),
(pseu\-do\-sca\-lar, vector),
(vector, positive-$C$-parity axial vector),
(scalar, vector),
(pseu\-do\-sca\-lar, pseu\-do\-sca\-lar),
(pseu\-do\-sca\-lar, positive-$C$-parity axial vector) and
(vector, negative-$C$-parity axial vector),
(pseu\-do\-sca\-lar, negative-$C$-parity axial vector),
(vector, vector in $P$ wave, total spin $S=0$).}
\label{vector}
\end{figure}
\begin{figure}[htbp]
\begin{center}
\begin{tabular}{c}
\includegraphics[height=479pt]
{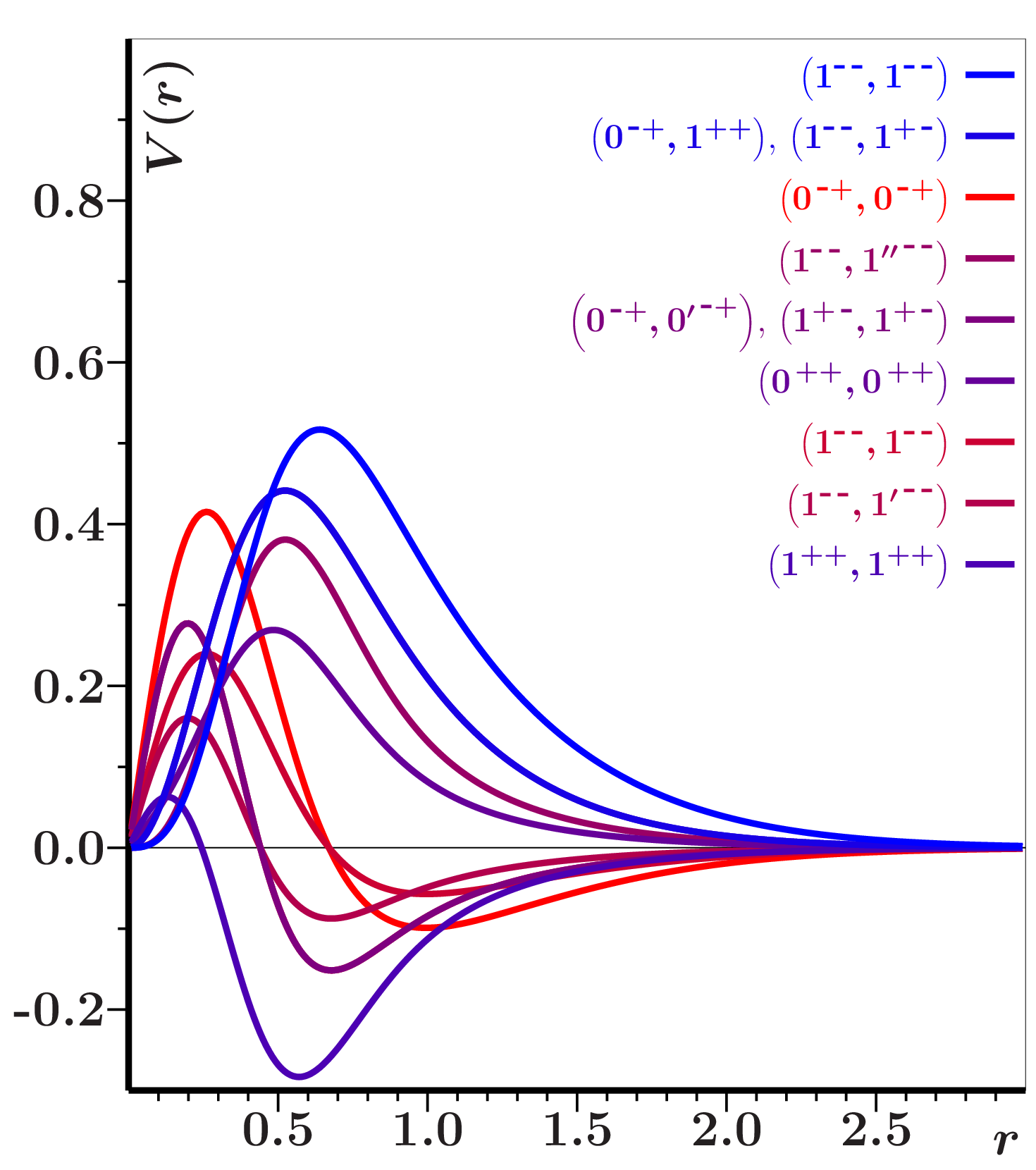}
\end{tabular}
\end{center}
\caption[]{\small
Transition potentials for the decay of scalar mesons
($r$ and $V(r)$ are dimensionless).
The curves are determined with expression (\ref{potentialMtoMM})
for the eleven three-meson vertices of Table~(\ref{Vertices}).\\
The curves,
from the curve with the highest maximum to the curve with the lowest maximum,
represent the scalar vertices with respectively
(vector, vector in $D$ wave, total spin $S=2$),
(pseu\-do\-sca\-lar, positive-$C$-parity axial vector) and
(vector, negative-$C$-parity axial vector),
(pseu\-do\-sca\-lar, pseu\-do\-sca\-lar),
(vector, vector (angular excitation)),
(pseu\-do\-sca\-lar, pseu\-do\-sca\-lar (radial excitation)) and
(negative-$C$-parity axial vector, negative-$C$-parity axial vector),
(scalar, scalar),
(vector, vector in $S$ wave, total spin $S=0$),
(vector, vector (radial excitation)),
(positive-$C$-parity axial vector, positive-$C$-parity axial vector).}
\label{scalar}
\end{figure}

In Figs.~\ref{vector} and \ref{scalar} are depicted similar results
for the vector and the scalar mesons respectively.
Is it reasonable in the case of pseudoscalar mesons to select two
dominant decay modes, in the case of vector mesons the situation is
less comfortable (see Fig.~\ref{vector})
and becomes even hopeless in the case of
scalar mesons (Fig.~\ref{scalar}).

So, in the next stage of the program of the multi-channel model of
\cite{PRD21p772,PRD27p1527} all possible decay channels should be
taken into account in order to decide afterwards which of them are
important for the description of a specific hadron.
The here presented method can supply us with the precise form
of the transition potentials.
Finally we must notice that the potentials of the
Figs.~\ref{pseudoscalar}, \ref{vector} and \ref{scalar},
exhibit a peaked structure.
In \cite{ZPC19p275} it is argued that such types of potentials
can very easily explain the radial hadron spectra, especially the large
level splittings between the ground states and the first radial
excitations with respect to the other mass differences.
\clearpage

\clearpage

\section*{Conclusions}

We have shown here that the method which was roughly outlined in
\cite{ZPC17p135}, can indeed be used
for the construction of transition potentials to be used in
coupled channel models \cite{PRD21p772,PRD27p1527}
for mesons (and baryons).
The procedure is still not unambiguous,
but, for a first inspection of the results, very suitable.
The next refinement could be, to take different quark
masses in the matrix elements (\ref{Mfinal}).
This leads to different transition
potentials for decay due to the creation of up and down quark pairs as
for decay due to the creation of strange quark pairs.
Another possibility is, to repeat the whole procedure for the $SO(3,2)$ wave
functions of \cite{THEFNYM7911},
which are probably more realistic than harmonic
oscillator wave functions.
\vspace{.3cm}

{\large\bf Acknowledgments}
\vspace{.3cm}

I thank Prof.~Dr.~C.~Dullemond for
the formulation of the program to be followed
and Dr. T.~A.~Rijken for many useful conversations.

Part of this work was included in the research program of the
{\it Stichting voor Fundamenteel Onderzoek der Materie} (F.O.M.)
with financial support from the
{\it Nederlandse Organizatie voor ZuiverWetenschappelijk Onderzoek} (Z.W.O.).


\clearpage


\begin{thebibliography}{14}
\bibitem{PRD17p3090}
E.~Eichten, K.~Gottfried, T.~Kinoshita, K.~D.~Lane and T.~M.~Yan,
{\it Charmonium:  1.  The Model},
Phys.\ Rev.\ D {\bf 17}, 3090 (1978)
[Erratum-ibid.\ D {\bf 21}, 313 (1980)].

E.~Eichten, K.~Gottfried, T.~Kinoshita, K.~D.~Lane and T.~M.~Yan,
{\it Charmonium: Comparison With Experiment},
Phys.\ Rev.\ D {\bf 21} (1980) 203.

\bibitem{PRL50p1181}
S.~Jacobs, K.~J.~Miller and M.~G.~Olsson
{\it Quarkonium bound states and coupling to hadrons},
Phys.\ Rev.\ Lett.\  {\bf 50}, 1181 (1983).

\bibitem{PRD29p110}
K.~Heikkil\"{a}, S.~Ono and N.~A.~T\"{o}rnqvist,
{\it Heavy $c\bar{c}$ and $b\bar{b}$ quarkonium states and unitarity effects},
Phys.\ Rev.\ D {\bf 29}, 110 (1984)
[Erratum-ibid.\ D {\bf 29}, 2136 (1984)].

\bibitem{KAZIMIERZ83p257}
C.~Dullemond, T.~A.~Rijken, E.~van Beveren and G.~Rupp,
{\it On The Influence Of Hadronic Decay On The Properties Of Hadrons},
in proceedings of {\it VIth Warsaw Symposium on Elementary Particle Physics,
Kazimierz, Poland, 30 May - 3 Jun 1983}, pp 257-262.

\bibitem{ZPC17p135}
E.~van Beveren,
{\it Recoupling matrix elements and decay},
Z.\ Phys.\ C {\bf 17}, 135 (1983)
[arXiv:hep-ph/0602248].

\bibitem{NPB10p521}
L.~Micu,
{\it Decay rates of meson resonances in a quark model},
Nucl.\ Phys.\ B {\bf 10}, 521 (1969).

R.~D.~Carlitz and M.~Kislinger,
{\it Regge amplitude arising from $SU(6)-W$ vertices},
Phys.\ Rev.\ D {\bf 2} (1970) 336.

\bibitem{PRD8p2223}
A.~Le Yaouanc, L.~Oliver, O.~P\`{e}ne and J.~C.~Raynal,
{\it Naive quark-pair-creation model of strong-interaction vertices},
Phys.\ Rev.\ D {\bf 8} (1973) 2223.

A.~Le Yaouanc, L.~Oliver, O.~P\`{e}ne and J.~C.~Raynal,
{\it Naive quark-pair-creation model and baryon decays},
Phys.\ Rev.\ D {\bf 9} (1974) 1415.

\bibitem{AP124p61}
M.~Chaichian and R.~K\"{o}gerler,
{\it Coupling constants and the nonrelativistic quark model with
charmonium potential},
Annals Phys.\ {\bf 124}, 61 (1980).

\bibitem{PRD21p772}
E.~van Beveren, C.~Dullemond, and G.~Rupp,
{\it Spectra and strong decays of $c\bar{c}$ and $b\bar{b}$ states},
Phys.\ Rev.\ D {\bf 21}, 772 (1980)
[Erratum-ibid.\ D {\bf 22}, 787 (1980)].

\bibitem{PRD27p1527}
E.~van Beveren, G.~Rupp, T.~A.~Rijken, and C.~Dullemond,
{\it Radial spectra and hadronic decay widths of light and heavy mesons},
Phys.\ Rev.\ D {\bf 27}, 1527 (1983).

\bibitem{Pilkuhn}
H.~Pilkuhn,
{\it The interaction of hadrons},
Amsterdam: North Holland publishing company, 1967

\bibitem{ZPC19p275}
E.~van Beveren, C.~Dullemond and T.~A.~Rijken,
{\it On the influence of hadronic decay on the properties of hadrons},
Z.\ Phys.\ C {\bf 19}, 275 (1983).

\bibitem{AP105p318}
C.~Dullemond and E.~van~Beveren,
{\it Confining potentials and Regge poles},
Ann.\ Phys.\ {\bf 105}, 318 (1977).

\bibitem{THEFNYM7911}
E.~van~Beveren, T.~A.~Rijken and C.~Dullemond,
{\it Quarks in an anti-De Sitter geometry},
Nijmegen-report THEF-NYM-79-11 (1979).
\end{thebibliography}
\end{document}